\documentclass[twocolumn,showpacs,amsmath,amssymb,groupedaddress,aps,prd]{revtex4-1}
\usepackage{graphicx}
\usepackage{dcolumn}
\usepackage{bm}
\usepackage{epsfig}
\usepackage{enumitem}

\begin{document}
\title{Constraining symmetron dark energy using atom interferometry}
\author{Sheng-wey Chiow}
\author{Nan Yu}
\email{nan.yu@jpl.nasa.gov}
\affiliation{Jet Propulsion Laboratory, California Institute of Technology, Pasadena, CA 91109}

\date{\today}

\begin{abstract}
Symmetron field is one of the promising candidates of dark energy scalar fields. 
In all viable candidate field theories, a screening mechanism is implemented to be consistent with existing tests of general relativity. 
The screening effect in the symmetron theory manifests its influence only to the thin outer layer of a bulk object, where inside a dense material the symmetry of the field is restored and no force exists. 
For pointlike particles such as atoms, the depth of screening is larger than the size of the particle, such that the screening mechanism is ineffective and the symmetron force is fully expressed on the atomic test particles. 
Extra force measurements using atom interferometry are thus much more sensitive than bulk mass based measurements, and indeed have placed the most stringent constraints on the parameters characterizing symmetron field in certain region. 
There is however no clear direct connection between the laboratory measurements and astrophysical observations, where the constraints are far separated by 10 orders of magnitude in the parameter space. 
In this paper, we present a closed-form expression for the symmetron acceleration of realistic atomic experiments. 
The expression is validated through numerical simulations for a terrestrial fifth-force experiment using atom interferometry. 
As a result, we show the connection of the atomic measurement constraints to the astrophysical ones.
We also estimate the attainable symmetron constraints from a previously proposed experiment in space intended for test of chameleon theory. 
The atomic constraints on the symmetron theory will be further improved by orders of magnitude.
\end{abstract}
\pacs{}
\maketitle

\section{introduction}
The accelerated expansion rate of the Universe is driven by dark energy, and the phenomena can be explained in the framework of scalar fields~\cite{DE1}. 
The interaction of the scalar field with normal matter should yield minute new forces of the strength of gravitational force.
Local scale experiments, however, have not yet detected forces on test objects apart from the four known forces, rendering the necessity that any dark energy scalar field must be environmentally dependent and thus the influence of dark energy would be greatly suppressed near dense material, which is known as the screening mechanism~\cite{DE2}. 
Scalar field parameters are in turn bounded by precise experiments on tests of the inverse-square law of gravity, bounds in the parametrized post Newtonian (PPN) metric, and tests of the equivalence principle, as summarized in Refs.~\cite{hui2009equivalence,sakstein2017tests} and references therein. 
Chameleon and symmetron theories are two possible scalar fields of simplicity and interest~\cite{DE2}.
The symmetron theory employs three parameters to achieve the screening mechanism~\cite{hinterbichler2010screening,hinterbichler2011symmetron,brax2015casimir}. 
Similar to the chameleon theories, small test particles do not suffer from the screening effect, favoring the approach of atom interferometric validation of theories~\cite{brax2016atomic,Burrage2015,BerkeleyScience,BerkeleyPRD,BerkeleyNaturePhys}. 
Unlike the chameleon theories, the constraints on symmetron by the atomic physics approach are reported to be narrow-ranged, and the connection to astrophysical observations has been remote thus far~\cite{sakstein2017tests,burrage2018tests}.

In this paper, we present a closed-form expression of symmetron acceleration for atomic test particles in realistic experimental settings.
The expression is validated via numerical simulations of the experiment detailed in Ref.~\cite{BerkeleyNaturePhys}.
With the results from Ref.~\cite{BerkeleyNaturePhys}, we obtained the constraints for symmetron to {\it allow direct comparison with astrophysical observations.}
Furthermore, the conceptual spaceborne experiment detailed in Ref.~\cite{chiow2018multiloop} is analyzed for symmetron sensitivity, showing that an improvement of 5 orders of magnitude can be achieved.
Finally, an ideal configuration for atomic tests of the symmetron model is discussed by utilizing the closed-form expression, which suggests to perform atom interferometers directly in the open-space vacuum at places such as the cislunar space, where a manned gateway facility will be established soon~\cite{LOPG,LOPG2,LOPG3}.

The article is organized in the following way. 
In Section~\ref{secII}, we briefly describe the symmetron model established in the astrophysical community. 
In Section~\ref{secIII}, we discuss the advantage of using cold atoms for detecting the symmetron field, followed by numerical calculation schemes for symmetron forces that an atom experiences. 
Then, we present result of constraints on symmetron parameters based on the experiment of Ref.~\cite{BerkeleyNaturePhys}, where closed-form expressions for estimating symmetron acceleration are also established. 
Moreover, in Section~\ref{secAstro}, we show constraints derived from laboratory atomic experiments in the same parameter range as from astrophysical observations.
Finally, in Section~\ref{secIV}, we conclude by proposing an atom interferometer experiment that will optimize the sensitivity to the symmetron field.

\section{The symmetron model}\label{secII}
Adapting the formalism from Refs.~\cite{hinterbichler2011symmetron,burrage2016constraining,sakstein2017tests,burrage2018tests}, the symmetron field $\phi$ exhibits a nonlinear self-interacting potential $\lambda \phi^4/4-\mu^2 \phi^2/2$, and a coupling to matter $\rho \phi^2 /M^2 /2$, where $\rho$ is the matter density and ($\lambda,\mu,M$) are parameters characterizing the behavior of the field.
With the effective potential $V_{\textrm{eff}}(\phi)=(\lambda \phi^4/4-\mu^2 \phi^2/2)+\rho \phi^2 /M^2 /2$, the equation of motion of a static symmetron field is
\begin{eqnarray}
\label{eq:sym}
\nabla^2 \phi &=& \frac{\partial V_{\textrm{eff}}}{\partial \phi} \nonumber \\
&=& \lambda \phi \left( \phi^2-\frac{\mu^2}{\lambda}\left(1-\frac{\rho}{\rho_*}\right)\right),
\end{eqnarray}
where $\rho_*\equiv M^2\mu^2$.
At ``equilibrium,'' by which $\nabla^2 \phi$ is defined to be 0 such that $\phi$ is space-invariant, $V_{\textrm{eff}}$ is minimized at $\phi=\pm\phi_e$ where
\begin{eqnarray}
\label{eq:phi_e}
\phi_e(\rho) &=&
\begin{cases} 
\frac{\mu}{\sqrt{\lambda}}\sqrt{1-\frac{\rho}{\rho_*}}, &\text{for $\rho<\rho_*$}\\
0, &\text{for $\rho\ge\rho_*$}
\end{cases}
\end{eqnarray}
and $\phi_e$ is called the vacuum expectation value (VEV).
Thus, in regions of low density, $\phi$ can settle to either $+\phi_e$ or $-\phi_e$ and the symmetry is said to be broken; on the other hand, in regions of high density, $\phi$ approaches zero and the symmetry is restored~\cite{sakstein2017tests}.

Expanding about $\phi=\phi_e$, the equation of motion reduces to $\nabla^2\left(\phi-\phi_e\right) \simeq m_s^2 (\phi-\phi_e)$, where $m_s$ is the mass of the field defined as 
\begin{eqnarray}
\label{eq:mass}
m_s^2 &\equiv& \left.\frac{\partial^2 V_{\textrm{eff}}}{\partial \phi^2}\right|_{\phi=\phi_e}\nonumber\\
            &=& \lambda \left( 3\phi_e^2-\frac{\mu^2}{\lambda}\left(1-\frac{\rho}{\rho_*}\right)\right)\nonumber\\
            &=& 
            \begin{cases}
            2\mu^2\left(1-\rho/\rho_*\right), & \text{for $\rho<\rho_*$}.\\
            \mu^2\left(\rho/\rho_*-1\right), & \text{for $\rho\geq\rho_*$}.
            \end{cases}
\end{eqnarray}
Thus, $\phi$ exponentially approaches or deviates from $\phi_e$ on the length scale of $m_s^{-1}$, the Compton wavelength of the field.

As a result, a metal chamber of thick walls (thickness $\gg m_s^{-1}\simeq\mu^{-1}(\rho/\rho_*)^{-1/2}$) isolates the internal field from the external, since the symmetry is restored in the walls that $\phi=0$, which allows determination of the internal field profile regardless of the mass distribution of the rest of the universe.
This property has been exploited in tests of dark energy theories using atom interferometers, where a vacuum chamber is required to maintain ultra high vacuum (UHV) for atom interferometer operations.

For a given field profile $\phi$, the resulting acceleration of an infinitesimal test particle is $-\phi\nabla\phi/M^2$~\cite{BerkeleyNaturePhys,hui2009equivalence}.
Since $\phi$ may be altered significantly after introducing a test object of finite size, the force experienced by the test object is effectively suppressed by a screening factor $\lambda_A$ ranging between 0 (totally screened) and 1 (completely unscreened), i.e., the measured acceleration is reduced to 
\begin{eqnarray}
\label{eq:symacc}
a_s&=&-\lambda_A\frac{\phi\nabla\phi}{M^2}.
\end{eqnarray}
This self-adjustment of the field and the force suppression manifest the screening mechanism needed in a feasible dark energy theory.

\section{constraints set by atom interferometers}\label{secIII}
Atom interferometers use individual atoms in a UHV environment as sensitive force probes.
The screening factor of an atom in vacuum, however, is not trivial.
On the one hand, the high density $\rho_N$ at nucleus makes an atom a screened test particle.
On the other hand, the small radius $R_N$ of a nucleus favors the unscreened scenario, in which the field doesn't reach $\phi_e=0$ inside the nucleus due to the small $R_N$~\cite{BerkeleyNaturePhys}.
Thus, for $m_N R_N<1$, the field is not modified much and the atom is considered unscreened; for $m_N R_N>1$, the field reaches $\phi_e=0$ inside the nucleus and the atom is screened.
Since $m_N^2\simeq \rho_N /M^2$, as a result, an atom behaves like a screened bulk for small $M$ and an unscreened point for large $M$.

The expression of the symmetron force between two spherical objects of radii $R_A$ and $R_B$ in vacuum of background density $\rho_o$ is derived in Ref.~\cite{burrage2016constraining}:
\begin{eqnarray}
\label{eq:symforce}
F(r)&=&4\pi Q_A Q_B \left(1+m_o R_B\right)\left(1+m_o r\right)\frac{e^{m_o (R_A-r)}}{r^2},\\
\text{where}&&\nonumber\\
\label{eq:symcharge}
Q_i&=&\left.\left(\phi_o-\phi_{i}\right)R_i\left(\frac{m_i R_i-\tanh{m_i R_i}}{m_i R_i+m_o R_i \tanh{m_i R_i}}\right)\right|_{i=A,B},
\end{eqnarray}
and $\phi_i$, $m_i$ are the VEV and the mass of the field in the corresponding medium (subscript $o$ for vacuum, $\phi_o=\phi_e(\rho_o)$) as defined in Eqs.~(\ref{eq:phi_e}) and (\ref{eq:mass}).
$Q_i$ is referred to as the symmetron charge, in analogy to the electric charge, that is responsible for the force between objects.
However, it is not applicable for more general cases, such as arbitrarily shaped objects, multiple objects, or inside an enclosed volume of vacuum.
Thus, numerical calculations of the field profile based on Eq.~(\ref{eq:sym}) and realistic experimental arrangements have to be conducted to estimate anticipated symmetron force for any $(\lambda,\mu,M)$.

\subsection{Numerical simulation}
We modify the software package detailed in Ref.~\cite{chiow2018multiloop} for the symmetron equation of motion Eq.~(\ref{eq:sym}).
In addition to changing the equation of motion from chameleon to symmetron, proper scaling of Eq.~(\ref{eq:sym}) in different regimes is essential for allowing a coverage of over 100 orders of magnitude in the parameter space.
It is because the difference of density between vacuum $\rho_o$ and the walls $\rho_w$ is about 17 orders of magnitude.
Moreover, $\rho_*$ is varied between simulations by at least 60 orders of magnitude, resulting in $1-\rho_w/\rho_*$ changing from a large negative number to almost 1.
Further, the evaluation of symmetron acceleration $a_s$ [Eq.~(\ref{eq:symacc})] requires both the field itself and its derivative, where the round-off error may severely limit the quality of $a_s$ particularly when the whole system is unscreened with a huge $\phi$ and tiny $\nabla \phi$.

In our simulation, Eq.~(\ref{eq:sym}) is transformed as follows. 
When the symmetry of the field is restored in the wall and in the source mass, the VEV in matter is $\phi_w=0$ and the field ranges between 0 and $\phi_o$ over the whole domain of simulation.
In this case, the field is scaled by $\phi_o^{-1}$ so that higher orders have reduced impact:
\begin{eqnarray}
\phi&=&\phi_o (1+\psi), \nonumber \\
\nabla^2 \psi &=& \frac{\rho-\rho_o}{M^2} +\left( 2\mu^2+\frac{\rho-3\rho_o}{M^2}\right)\psi\nonumber\\
&&+3\lambda\phi_o^2\psi^2+\lambda\phi_o^2 \psi^3.
\end{eqnarray}
When the system is unscreened in matter, $\phi_w^2=\mu^2 (1-\rho_w/\rho_*)/\lambda$ [Eq.~(\ref{eq:phi_e})], the field is close to $\mu/\lambda^{1/2}$ everywhere.
The equation of motion is solved by offsetting the solution by $\phi_\text{off}=\mu/\lambda^{1/2}$:
\begin{eqnarray}
\phi&=&\phi_\text{off}+\psi, \nonumber \\
\nabla^2 \psi &=& \frac{\rho}{M^2} \phi_\text{off}+\left( 2\mu^2+\frac{\rho}{M^2}\right)\psi \nonumber\\
&&+\lambda \left( 3\phi_\text{off} \psi^2+\psi^3 \right).
\end{eqnarray}
The profile of the VEV, i.e., $\phi_e(\rho(\vec{r}))$, of the simulation domain is used as the initial guess for the solver after respective scaling or offsetting.

To test the robustness of the solution, we check the consistency of solutions by multiplying the initial guess by a factor of 10, 100, and 1000.
We also check the solution stability against small variations in $\mu$ or $M$, e.g., by a fraction of $10^{-4}$.
Only solutions passing these checks will be used.

\subsection{The screening factor}
The screening factor $\lambda_A$ characterizes how the scalar field is modified by the presence of the test object and thus leading to a weaker symmetron force than that would have been measured by an ideal test particle, which presumably does not alter the scalar field.
The concept of screening factor is useful when the scalar field is to be probed by a small object, such as an atom.
In Ref.~\cite{BerkeleyNaturePhys}, the screening factor of an atom, particularly its nucleus, is estimated by taking the ratio of the symmetron charge of the nucleus $Q_N$, which is calculated in the scenario of two spherical objects [Eq.~(\ref{eq:symcharge})], to its value when the nucleus is unscreened.

The approach of using the symmetron charge calculated with two spherical objects and taking the ratio of itself in the unscreened limit is well motivated, but not justified.
Moreover, applying an additional factor to simulation results, which base only on the governing equation of motion, may constitute a loophole in testing the theory itself.
In this section, we demonstrate a numerical method to validate the symmetron charge approach, and at the same time to provide symmetron force estimates relying purely on simulation of the symmetron equation of motion of Eq.~(\ref{eq:sym}).

The screening charge approach requires finding the hypothetical unscreened symmetron charge $Q_{N,\text{u}}$ of the nucleus. 
The nucleus is unscreened when the symmetron Compton wavelength $m_N^{-1}$ is larger than the nuclear size such that the nucleus has no impact on the field profile. 
Therefore, the hypothetical unscreened charge is defined to be the leading term of $Q_N$ in Eq.~(\ref{eq:symcharge}) in the limit of $m_N R_N \ll 1$: $Q_{N,\text{u}}\equiv\left(\phi_o-\phi_N\right)R_N\frac{m_N^2 R_N^2}{3(1+m_o R_N)}$, even though $m_N R_N$ can be large for the parameters under consideration.
Thus, the screening factor of the nucleus is
\begin{eqnarray}
\label{eq:screeningfactor}
\lambda_A & \equiv & \frac{Q_N}{Q_{N,\text{u}}} \nonumber\\
&=&\frac{3(1+m_o R_N)}{m_N^2 R_N^2}\frac{m_N R_N-\tanh{m_N R_N}}{m_N R_N+m_o R_N \tanh{m_N R_N}}\\
&\simeq&
\begin{cases}
1-\dfrac{2}{5} \dfrac{\rho_N R_N^2}{M^2},\  \text{for $m_N R_N \ll1$ (unscreened)}\\[+3mm]
\dfrac{ 3(1+\sqrt{2}\mu R_N)}{R_N^2 \rho_N}M^2,\ \text{for $m_N R_N \gg 1$ (screened)}
\end{cases},\nonumber
\end{eqnarray}
assuming $\rho_N>\rho_*$.
Since $R_N$ is very small, e.g., $3.2\times 10^{-8}$~eV$^{-1}$ for Cs (one of the atomic species used in symmetron experiments), $\mu R_N$ is much less than 1 for $\mu<1$~eV, such that $\lambda_A$ depends largely only on $M$ and independent of $\lambda$ and $\mu$.
In the discussions to follow, therefore, we will exam the screening dependence on $M$.

We simulate the screening factor by computing the screened symmetron acceleration experienced by a nucleus in a symmetron field of constant gradient.
Specifically, it is a simulation of Eq.~(\ref{eq:sym}) in the cylindrical coordinates $(r,z)$ of a sphere at $(0,0)$ of density $\rho_N$ and radius $R_N$ immersed in vacuum of density $\rho_o$, with a spherical domain of radius $R_\text{d} \sim 10R_N$.
The boundary condition of $\phi$ is set to $\phi_b(1+\eta z/R_\text{d} )$, where $\eta\sim 0.1$ and $\phi_b$ is typically chosen to be $\phi_o$, the VEV corresponding to the background vacuum density.
This boundary condition mimics the situation of a constant gradient $\eta\phi_b/R_\text{d}$ in the $z$-direction on top of a constant background $\phi_b$.
After finding the solution of $\phi(r,z)$, the acceleration experienced by the nucleus is calculated as
\begin{eqnarray}
\label{eq:nuclearacc}
a_N&=&\frac{\displaystyle\iint -\dfrac{\rho_N \phi(r,z)\, \partial_z \phi(r,z)}{M^2} 2\pi r\, dr\, dz}{\displaystyle\iint \rho_N\ 2\pi r\, dr\, dz}\nonumber\\
&=&-\frac{\displaystyle\iint r \phi(r,z)\, \partial_z \phi(r,z)\,  dr\, dz}{M^2 \displaystyle\iint r\, dr\, dz},
\end{eqnarray}
where the integrals are evaluated over the region of the semicircle, $r^2+z^2\le R_N^2, r\ge0$, and the denominator is evaluated numerically as a check for errors in meshing or in area computation, despite that its closed-form is well-known.
On the other hand, the unscreened acceleration that would have been measured by an ideal test particle is $-\phi\nabla\phi/M^2=-\eta\phi_b^2/R_\text{d}M^2$.
The screening factor is then the ratio of $a_N$ in Eq.~(\ref{eq:nuclearacc}) to that of the ideal test particle:
\begin{eqnarray}
\label{eq:screeningfactorSIM}
\lambda_A&=&\frac{a_N}{-\eta\phi_b^2/R_\text{d}M^2}\nonumber\\
&=&\frac{\displaystyle\iint r \phi(r,z)\, \partial_z \phi(r,z)\,  dr\, dz}{\frac{\eta}{R_\text{d}}\phi_b^2 \displaystyle\iint r\, dr\, dz}.
\end{eqnarray}

\begin{figure}
\centering
\includegraphics[width=0.45\textwidth]{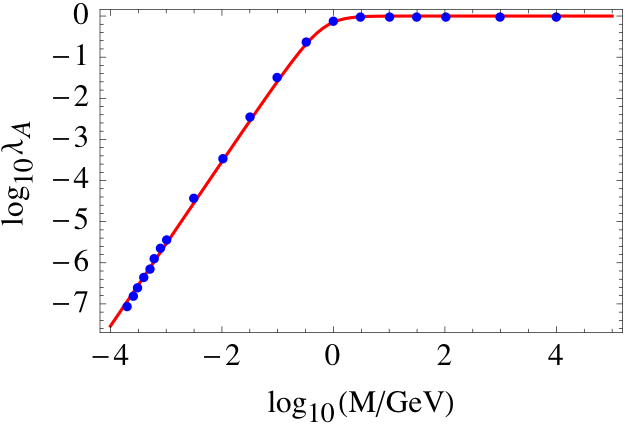}
\caption{Comparison of screening factor $\lambda_A$ calculation.
Red solid curve: $\lambda_A$ from two-sphere theory [Eq.~(\ref{eq:screeningfactor})].
Blue points: $\lambda_A$ from the numerical simulation [Eq.~(\ref{eq:screeningfactorSIM})].
Note that there is no fitting to Eq.~(\ref{eq:screeningfactor}).
Parameters used are $\rho_o=6.6\times 10^{-17}\text{~g}/\text{cm}^3$, $\rho_N=2.3\times 10^{14}\text{~g}/\text{cm}^3$, $R_N=6.38$~fm, and $\mu=10^{-4}$~eV.
Choices of $\lambda$, $\phi_b$, $\eta$, or $R_b\ge 10R_N$ have no noticeable influence on $\lambda_A$.
}
\label{fig:screeningfactor}
\end{figure}

Figure~\ref{fig:screeningfactor} shows a comparison of $\lambda_A$ calculated using Eq.~(\ref{eq:screeningfactor}) (solid curve) and  $\lambda_A$ simulated using Eq.~(\ref{eq:screeningfactorSIM}) (scattered points) for $\mu=10^{-4}$~eV, $\rho_o=6.6\times 10^{-17}\text{~g}/\text{cm}^3$, and $\rho_N=2.3\times 10^{14}\text{~g}/\text{cm}^3$.
We find that the simulated $\lambda_A$ is insensitive to $\eta$ and $\phi_b$, as expected.
Clearly, $\lambda_A$ from the simulation of a nucleus in a symmetron field of constant gradient agrees with the symmetron charge approach derived from two spherical objects, except at small $M$.
The simulation data exist only for $M>\sqrt{\rho_o}/\mu=1.68\times10^{-4}$~GeV; for smaller $M$, $\rho_*$ is smaller than $\rho_o$, implying that symmetry is restored even in vacuum such that $\phi=0$ everywhere [Eq.~(\ref{eq:phi_e})], and that no symmetron force at all.
However, $\lambda_A$ from Eq.~(\ref{eq:screeningfactor}) exists for all $M$.
Nevertheless, since there is no symmetron force for $M< \sqrt{\rho_o}/\mu$, the screening factor is not defined and not relevant.

The numerical confirmation of Eq.~(\ref{eq:screeningfactor}) serves three purposes here.
First, even though Eq.~(\ref{eq:symcharge}) is derived with approximations such that no exact solution of Eq.~(\ref{eq:sym}) is obtained, we verify that the screening factor calculated is valid for a large parameter range.
Second, the software package for solving Eq.~(\ref{eq:sym}) is also validated, since it is unlikely that both approaches are inadequate in the same way over a large span of parameter $M$.
Third, the approach of considering only experimental configurations with ideal test particles and then applying the corresponding screening factor afterwards can now be regarded as simulations based only on the master equation of Eq.~(\ref{eq:sym}), as oppose to needing the oversight of the theoretically motivated screening factor.

\subsection{Terrestrial experiment: symmetron acceleration on atomic test particles}
The symmetron force exerted on atomic test particles depends on $(\lambda,\mu,M)$ and the geometry between atoms and the source mass.
Due to vastly different length scales of the atomic nuclei and the vacuum chamber, it is not practical to simulate Eq.~(\ref{eq:sym}) including both atoms and the source mass.
Fortunately, as we find from the screening factor simulations, the extent of influence of a nucleus to the symmetron field is limited to few $R_N$.
The calculation of symmetron acceleration of an atom can thus be divided into two parts.
First, given $(\lambda,\mu,M)$, Eq.~(\ref{eq:sym}) is solved for the symmetron field profile $\phi$ based on actual experimental arrangement, i.e., the profile of matter density $\rho$, while ignoring the presence of cold atoms.
Second, the response of atoms to $\phi$ is altered by the screening factor $\lambda_A$, which depends mostly on $M$ and can be obtained via Eq.~(\ref{eq:screeningfactor}) or simulation as laid out in the previous section.

\begin{figure}
\centering
\includegraphics[width=0.45\textwidth]{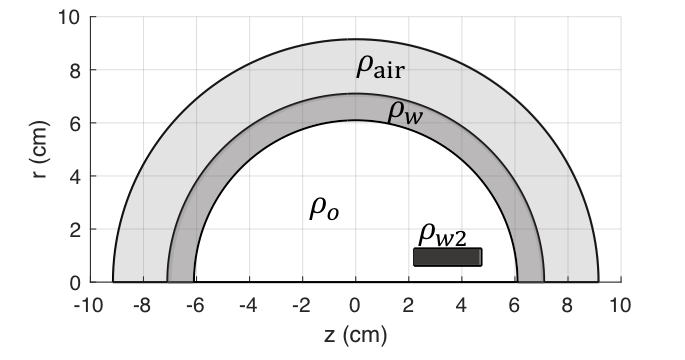}
\caption{Configuration for simulating the experiment of Ref.~\cite{BerkeleyNaturePhys} in the cylindrical coordinates.
There are four regions of different densities: UHV inside the shell: $\rho_o=6.6\times10^{-17}$~g/cm$^3$, the spherical stainless steel shell: $\rho_w=7$~g/cm$^3$, the cylindrical tungsten source mass with a through hole: $\rho_{w2}=19.3$~g/cm$^3$, and the ambient air outside the vacuum chamber: $\rho_\text{air}=1.225\times10^{-3}$~g/cm$^3$.
}
\label{fig:berkeley}
\end{figure}
Now, we show how one can obtain the symmetron constraints from the experimental results of Ref.~\cite{BerkeleyNaturePhys}, where a constraint of $a_\text{bound}<49$~nm/s$^2$ is placed on unknown forces.
Although the experimental setup has no axial symmetry, a cylindrical symmetric configuration is used (depicted in Fig.~\ref{fig:berkeley}) for reducing complexity and computing resources while capturing major features.
Specifically, we consider a tungsten cylinder with an axial thru-hole placed on axis but off-center inside a steel spherical shell, where the background gas pressure is $6\times10^{-10}$~torr inside and $760$~torr outside the shell for UHV and the ambient air, respectively.
To account for the difference in geometry, particularly the slot in the tungsten cylinder in Ref.~\cite{BerkeleyNaturePhys}, the radius of the through hole is adjusted to best match the field profile $\lambda \phi \nabla\phi$ shown in Ref.~\cite{BerkeleyNaturePhys}.
We find that a radius of $0.615$~cm yields a satisfactory result.
While the best effort is made to represent the experimental configuration for simulation, we note that an uncertainty factor of order one will not significantly undermine the conclusion for such a null measurement.
Refined three-dimensional simulations can be conducted when there are parameter regions of particular interest, including near-overlap exclusion regions from distinct measurement types, occurrence of non-zero fifth force measurements, etc.

As atoms are tossed up and freely falling down in the terrestrial experiment, the symmetron acceleration is not uniform during the measurement.
We use the triangular response function of an atom interferometer to time dependent accelerations to calculate the time-weighted acceleration~\cite{geiger2011detecting}.

\subsection{Closed-form expression for estimating symmetron acceleration}\label{iiid}
\begin{figure}
\centering
\includegraphics[width=0.45\textwidth]{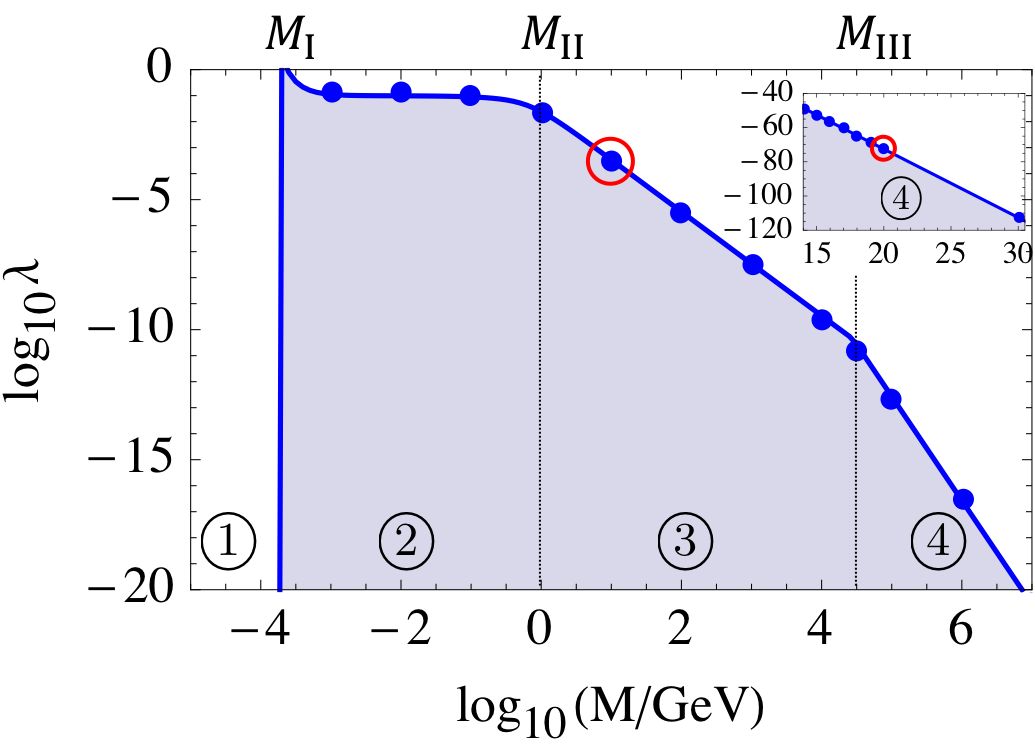}
\caption{$M$-$\lambda$ exclusion plot for $\mu=10^{-4}$~eV.
Inset shows the region of large $M$.
Solid dots represent simulation results.
Solid curve shows the fit of Eq.~(\ref{eq:as1}) using two circled data points for an overall scaling factor and $M_\text{III}$.
Regions of different $M$-$\lambda$ dependences are indicated, as discussed in text.
}
\label{fig:lambdaM4}
\end{figure}
An analytic expression of $a_s$ as a function of $(\lambda,\mu,M)$ for a realistic experiment is not generally available, as opposed to the symmetron force given in Eq.~(\ref{eq:symforce}) for two spheres.
Here we describe a recipe for obtaining an expression of $a_s$ for a specific experiment, which agrees with simulation results fairly well over several orders of magnitude in all $(\lambda,\mu,M)$.
This procedure will significantly reduce the simulation effort and help optimize the design of future experiments.

The recipe was inspired by an argument in Ref.~\cite{BerkeleyNaturePhys}, which is briefly summarized as follows.
Since $\phi\sim \mu/\sqrt{\lambda}$ [Eq.~(\ref{eq:phi_e})], $\phi\nabla\phi\sim \mu^2/\lambda$ such that $\lambda\phi\nabla\phi$ should be invariant over $M$ for a fixed $\mu$.
This invariance was indeed demonstrated in their work, and was used to obtain the exclusion region based on their experimental bound on unknown acceleration $a_\text{bound}<49$~nm/s$^2$: $a_\text{bound}\propto \phi\nabla\phi/M^2\sim \mu^2/\lambda M^2$, thus $\lambda\sim M^{-2}$ for generating a fixed symmetron acceleration.
Once $\lambda=\lambda_\text{sim}$ is found by simulation to yield a symmetron acceleration of magnitude $a_\text{bound}$ for a chosen pair of $(\mu,M)=(\mu_\text{sim},M_\text{sim})$, the exclusion region for $\mu=\mu_\text{sim}$ is bounded by the curve $\lambda= \lambda_\text{sim} M_\text{sim}^{2}/M^2$, while the effects of $\lambda_A$ and other considerations have to be implemented by hand afterwards.

We establish the closed-form symmetron acceleration as follows.
From Eq.~(\ref{eq:symacc}), $a_s$ is a product of a symmetron field $\phi$ and a gradient $\nabla \phi$.
Since the measurement is conducted in vacuum, and the relevant scale of field difference is the VEVs between vacuum and the source mass, we anticipate 
\begin{eqnarray}
\label{eq:as1}
a_s&\propto&-\lambda_A\frac{\phi_o (\phi_o-\phi_w)}{M^2},
\end{eqnarray}
where $\phi_w$ is the VEV inside the source mass.
There are three values of $M$ that are of particular importance for the exclusion region of a given $\mu$, as illustrated in Fig.~\ref{fig:lambdaM4}.
First, when $M< \sqrt{\rho_o}/\mu\equiv M_\text{I}$ (Region~{\large \textcircled{\small 1}}), the symmetry is restored in vacuum ($\phi_o=0$) such that there is no symmetron force and the parameters are unbounded.
Second, when $M_\text{I}< M< M_\text{II}$ (Region~{\large \textcircled{\small 2}}), where $m_N R_N=1$ is satisfied at $M=M_\text{II}\simeq \sqrt{\rho_N}R_N$, the nucleus is screened with $\lambda_A\propto M^2$ [Eq.~(\ref{eq:screeningfactor})].
Combining with the above discussed acceleration dependence of $a_s\propto\phi\nabla\phi/M^2\sim \mu^2/\lambda M^2$, the symmetron acceleration measured by an atom is in fact independent of $M$ in Region~{\large \textcircled{\small 2}}: $a_s=-\lambda_A\phi\nabla\phi/M^2\sim \mu^2/\lambda$.
For $M>M_\text{II}$ (Region~{\large \textcircled{\small 3}}), $\lambda_A\simeq 1$ and $a_s\sim \mu^2/\lambda M^2$, as argued in Ref.~\cite{BerkeleyNaturePhys} and summarized earlier.
A fixed $a_s$ results in $\lambda\propto M^{-2}$ for a given $\mu$.
Third, when $M\ge \sqrt{\rho_w}/ \mu\equiv M_\text{III}$ (Region~{\large \textcircled{\small 4}}), where $\rho_w$ is the density of the source mass or the vacuum chamber, all materials in the experiment are unscreened [Eq.~(\ref{eq:phi_e})].
While this region was considered unconstrainable~\cite{burrage2018tests,BerkeleyNaturePhys}, we find that the symmetron force is present albeit suppressed.
When unscreened, $\rho<\rho_*$, and $\phi_e^2 = \frac{\mu^2}{\lambda}\left(1-\frac{\rho}{\rho_*}\right)$ [Eq.~(\ref{eq:phi_e})], such that
\begin{eqnarray}
a_s &\propto&-\dfrac{\phi_o\left(\phi_o-\phi_w\right)}{M^2}\nonumber \\
&\simeq&-\dfrac{\mu^2}{\lambda M^2}\left(1-\frac{\rho_o}{2\rho_*}\right)\left(\frac{\rho_w-\rho_o}{2\rho_*}\right)\nonumber \\
&\simeq&-\dfrac{1}{\lambda M^4}\left(\frac{\rho_w-\rho_o}{2}\right)\nonumber \\
&\propto&\left(\lambda M^4\right)^{-1}.
\end{eqnarray}
Thus, for a fixed $a_s$ at a given $\mu$, $\lambda\propto M^{-4}$ in Region~{\large \textcircled{\small 4}}.

In short, for a given $\mu$, the symmetron acceleration $a_s\simeq-\lambda_A\phi_o (\phi_o-\phi_w)/M^2$ is the same when: $\lambda$ unbounded for $M<M_\text{I}$ (Region~{\large \textcircled{\small 1}});  $\lambda\propto M^0$ for $M_\text{I}<M<M_\text{II}$ (Region~{\large \textcircled{\small 2}});  $\lambda\propto M^{-2}$ for $M_\text{II}<M<M_\text{III}$ (Region~{\large \textcircled{\small 3}}); and $\lambda\propto M^{-4}$ for $M_\text{III}<M$ (Region~{\large \textcircled{\small 4}}).
Figure~\ref{fig:lambdaM4} shows a comparison of simulation results and the above expression for $\mu=10^{-4}$~eV, where Regions {\large \textcircled{\small 1}}-{\large \textcircled{\small 3}} agree with previous publications~\cite{BerkeleyNaturePhys,burrage2016constraining} while Region~{\large \textcircled{\small 4}} is new result from this study.
To have a quantitative comparison, Eq.~(\ref{eq:as1}) is fit for an overall scaling factor and $M_\text{III}$ with two simulation data points (circled in Fig.~\ref{fig:lambdaM4}) at $M=10^{20}$~GeV and $M=10$~GeV where $\lambda_A\simeq 1$ for both $M$ values.
$M_\text{I}$ and $M_\text{II}$ are determined solely by the vacuum density $\rho_o$ and properties of Cs nucleus.
Clearly, Eq.~(\ref{eq:as1}) describes the boundary of the exclusion region very well using only two simulation runs.
Both the simulation and the theoretical estimate [Eq.~(\ref{eq:as1})] support that the exclusion of $M$ goes well beyond $M_\text{III}$ set by the source mass~\footnote{The characteristic density $\rho_w$ corresponding to the fitted $M_\text{III}$ for $\mu=10^{-4}$~eV (Fig.~\ref{fig:lambdaM4}) is about 1.8~g/cm$^3$, which is smaller than the density of the source mass (19.3~g/cm$^3$) and the density of the shell (7~g/cm$^3$), suggesting that the system is unscreened at lower $M$ than that determined by the material density.
It is understood that the symmetron field can tunnel through the wall of finite thickness when the material is nearly unscreened.
Nevertheless, the difference in $\log_{10}M_\text{III}$ is only 0.5 between 1.8~g/cm$^3$ and 19.3~g/cm$^3$, which is negligible compared to the span of $M$ under discussion.
}, and the validity of $\lambda\sim M^{-4}$ is currently verified up to the simulation capability at about $M=10^{30}$~GeV.
Note that the smooth transition between Regions shows that different rescaling of Eq.~(\ref{eq:sym}) for numerical simulation are consistent and that the exact distinction between rescaling methods is not critical.

\begin{figure}
\centering
\includegraphics[width=0.45\textwidth]{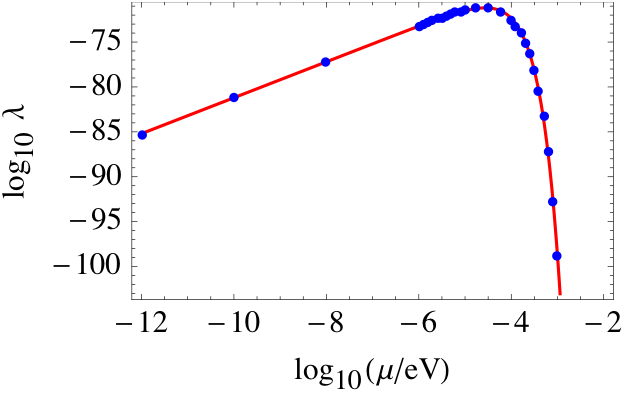}
\caption{$\mu$-$\lambda$ dependence at $M=10^{20}$~GeV.
Red curve shows the fit of data points to $\mu^2 e^{-\sqrt{2}\mu r_*}$.
The fit $r_*=52\times10^3$~eV$^{-1}$ corresponds to 1.0~cm.
}
\label{fig:lambdamu20}
\end{figure}

The simulation is extended to other values of $\mu$.
At $M=10^{20}$~GeV, $\mu$-$\lambda$ dependence is established, as shown in Fig.~\ref{fig:lambdamu20}.
We find the data are well described by $\lambda\sim\mu^2 e^{-\sqrt{2} \mu r_*}$, where $r_*$ is a characteristic length scale.
This relation is in agreement with the two spherical body scenario described by Eq.~(\ref{eq:symforce}), where $m_o\simeq \sqrt{2}\mu$ when $M$ is large [Eq.~(\ref{eq:mass})].
Particularly, the fit of $r_*\simeq 1$~cm is indeed the distance of the atoms to the source mass.
This geometrical dependence is not captured in Eq.~(\ref{eq:as1}), which is motivated from the local field perspective [Eq.~(\ref{eq:symacc})].
Thus, a more precise estimate of the symmetron acceleration in the experiment is
\begin{eqnarray}
\label{eq:as2}
a_s&\propto&-\lambda_A\frac{\phi_o (\phi_o-\phi_w)}{M^2} m_o^2 e^{-m_o r_*},
\end{eqnarray}
where $r_*$ is to be determined by fitting the $\mu$-$\lambda$ dependence at large $M$.
The introduction of the additional $m_o$ dependence yields the sharp increase near $M_\text{I}$ in Fig.~\ref{fig:lambdaM4}, which will be discussed later in the section.

In Regions {\large \textcircled{\small 2}} and {\large \textcircled{\small 3}} where the symmetry is restored in the walls ($\phi_w=0$), another constraint exists.
Consider the field inside a thick spherical shell of inner radius $R_w$, where $\phi=\phi_w=0$ inside the sufficiently thick wall.
Deviating from the trivial solution of universal $\phi=0$, a perturbative field $\delta \phi$ satisfies (from Eq.~(\ref{eq:sym}))
\begin{eqnarray}
\nabla^2 \delta\phi &\simeq& -\mu^2 \left(1-\frac{\rho_o}{\rho_*}\right) \delta \phi=-\lambda \phi_o^2 \delta \phi,
\end{eqnarray}
and the boundary conditions for the radial component in the spherical coordinates are $\partial_r \delta \phi=0$ at the origin and $\delta \phi=0$ at $r=R_w$.
The solution is an even-order spherical Bessel function $j_n(\sqrt{\lambda}\phi_o r)$, and $j_n(\sqrt{\lambda}\phi_o R_w)=0$.
The lowest allowed $\sqrt{\lambda}\phi_o$ is $\sqrt{\lambda}\phi_o R_w=\pi$, or equivalently $\mu\gtrsim \pi /R_w\equiv\mu_w$ ($\mu_w\simeq 10^{-5}$~eV for $R_w=6.1$ cm).
In other words, for $\mu < \mu_w$ the symmetron field is zero inside the shell and no symmetron force.
For $\mu \ge \mu_w$, the symmetron field is not zero if the field outside the shell is not zero, i.e., $\phi_\text{air}>0$ or equivalently $M>\sqrt{\rho_\text{air}}/\mu\equiv M_\text{III}^*$.
In this case, since the field is not separated by the wall, it is in the domain of Region~{\large \textcircled{\small 4}}.
With a small source mass inside the shell as in the real experiment, $\mu_w$ will change slightly and the transition from zero to nonvanishing field will be smooth.
\begin{figure}
\centering
\includegraphics[width=0.45\textwidth]{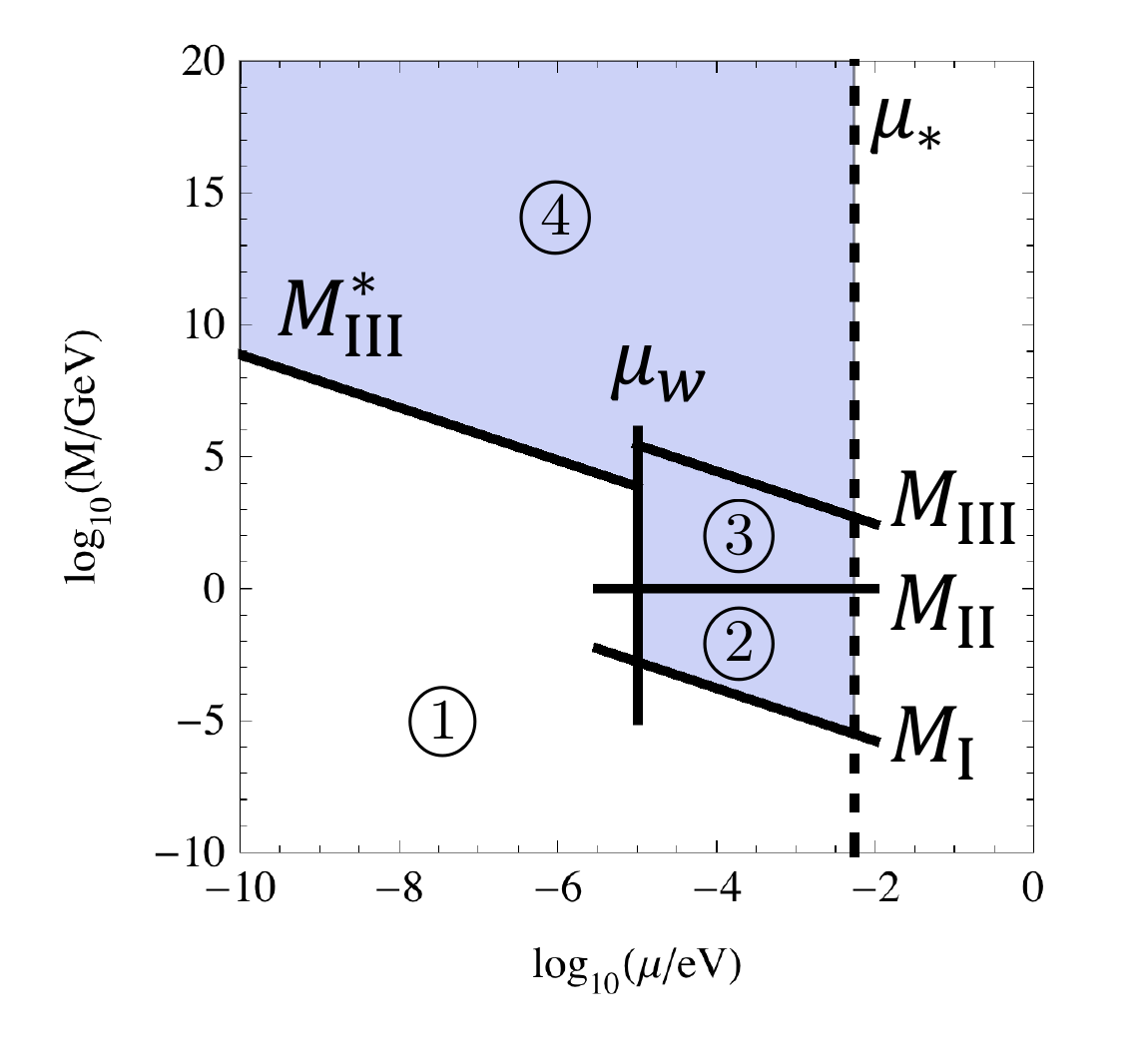}
\caption{Characteristic boundaries of the exclusion region of symmetron parameters $(\lambda,\mu,M)$ (shaded area).
For a given $\mu$, no sensitivity in Region~{\large \textcircled{\small 1}}, and a given symmetron acceleration can be reached by the scaling rules: $\lambda\propto M^0$ in Region~{\large \textcircled{\small 2}}, $\lambda\propto M^{-2}$ in Region~{\large \textcircled{\small 3}}, and $\lambda\propto M^{-4}$ in Region~{\large \textcircled{\small 4}}.
}
\label{fig:bounds}
\end{figure}

Note that the attempt to establish acceleration scaling versus $(\lambda,\mu,M)$ is also reported in Ref.~\cite{upadhye2013symmetron} and Ref.~\cite{sabulsky2018experiment}, in addition to Ref.~\cite{BerkeleyNaturePhys}.
Ref.~\cite{upadhye2013symmetron} uses the one-dimensional plane-parallel (1Dpp) approximation for the symmetron field between two parallel plates, and Ref.~\cite{sabulsky2018experiment} introduces a fitting factor of order 1 to extend the 1Dpp approximation to the interior of a vacuum chamber.
Equation~(\ref{eq:as2}), on the contrary, is applicable to a wide range of $(\lambda,\mu,M)$ with fixed geometric parameters.
It also allows exclusion in Region~{\large \textcircled{\small 4}}.

To summarize, the constraints of symmetron parameters $(\lambda,\mu,M)$ are characterized by several bounds associated with experimental parameters, as depicted in Fig.~\ref{fig:bounds}.
$M_\text{I}= \sqrt{\rho_o}/\mu$, below which the symmetry of the field is restored in vacuum and the field is zero everywhere.
$M_\text{II}=\sqrt{\rho_N}R_N$, below which the nucleus of the atomic test particle is screened with a screening factor $\lambda_A\propto M^{-2}$.
$M_\text{III}=\sqrt{\rho_w}/ \mu$, below which the symmetry is restored in the vacuum chamber walls, and above which the whole system is unscreened.
$M_\text{III}^*=\sqrt{\rho_\text{air}}/ \mu$, below which the symmetry is restored in air and $\phi_\text{air}=0$.
$\mu_w=\pi /R_w$, below which the field is zero inside the enclosure if the field outside is zero.
$\mu_*\sim280/r_*$, at which the field is about $e^{-400}$ of its peak value for a given $M$.

\begin{figure}
\centering
\includegraphics[width=0.46\textwidth]{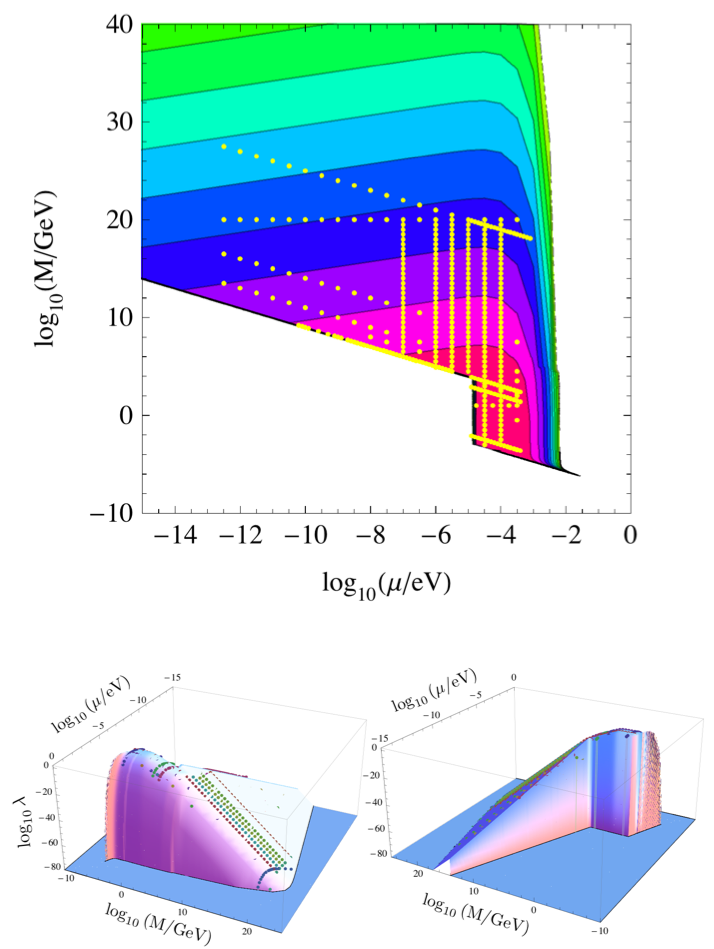}
\caption{Comparison of simulation results and the closed-form expression of Eq.~(\ref{eq:as2}).
Top: Yellow points are simulated parameters, and the contours (in steps of 20 dB) are $\log_{10}\lambda$ calculated based on the recipe.
Bottom: The surface of $(\lambda,\mu,M)$ is calculated from the recipe, on which the symmetron acceleration is the same, and the simulated points are plotted in two views.
The simulation parameter sets are chosen to verify the dependence of $\lambda$ when $M$, $\mu$, or the product $M\mu$ is kept constant in different regimes.
}
\label{fig:contour}
\end{figure}
The procedure for finding the exclusion region for a specific experimental symmetron acceleration upper bound is as follows.
First, determine the characteristic length scale $r_*$. 
It can be found by fitting simulation results at large $M$ for different $\mu$, as in Fig.~\ref{fig:lambdamu20}, where the screening factor is essentially 1 and the system is unscreened.
Or, one can take the approximate distance of the atomic cloud to the source mass as $r_*$.
Second, identify the densities of vacuum $\rho_o$, the wall $\rho_w$, the ambient air $\rho_\text{air}$, and the nuclear properties $\rho_N, R_N$.
Third, identify the characteristic size of the enclosed vacuum $R_w$.
Finally, based on the simulation result of one parameter set $(\lambda,\mu,M)$, apply the $M$-$\lambda$ scaling rules in Regions {\large \textcircled{\small 1}}-{\large \textcircled{\small 4}} and $\lambda\sim\mu^2 e^{-\sqrt{2} \mu r_*}$ for fixed $M$ [or equivalently Eq.~(\ref{eq:as2})] while respecting the bounds of $M_\text{III}^*$ and $\mu_w$ laid out in Fig.~\ref{fig:bounds}.
Figure~\ref{fig:contour} shows the comparison of simulation results at various $(\lambda,\mu,M)$ versus the region described by the above procedure, which is a realization of Eq.~(\ref{eq:as2}) with modifications imposed by $\mu_w$ and $M_\text{III}^*$.
Clearly the simulation results reside on the surface defined by the recipe except very near the edges of $\mu_w$ and of $M_\text{III}^*$.
Note that there is a lip around $\mu\sim10^{-2}$~eV in the surface plot of the exclusion region, which corresponds to the sharp increase of $\lambda$ near $\log_{10}(\mu/\text{eV})=-4$ in Fig.~\ref{fig:lambdaM4}.
It is because $m_o\sim 0$ when $M\gtrsim M_\text{I}$ ($\rho_o\lesssim \rho_*$) such that the exponential factor of $e^{-m_0 r_*}$ in Eq.~(\ref{eq:as2}) approaches unity and that the sensitivity is recovered in a narrow range of $M$ bounded by the $m_o^2$ factor.

\subsection{Connection to astrophysics}\label{secAstro}
\begin{figure}
\centering
\includegraphics[width=0.45\textwidth]{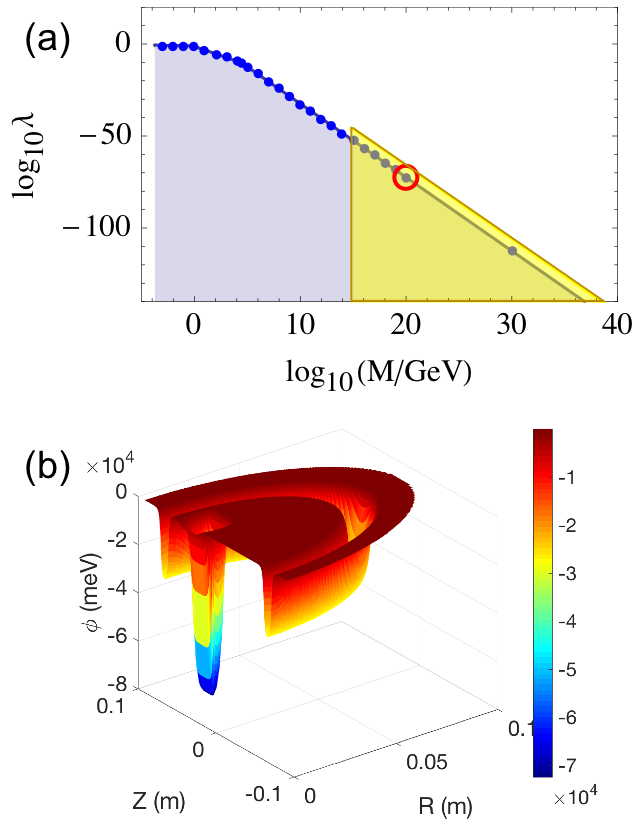}
\caption{Astrophysical constraints and atomic physics exclusion at $\mu=10^{-4}$~eV.
(a) Yellow region is the excluded region from astrophysical observations~\cite{burrage2018tests}.
As in Fig.~\ref{fig:lambdaM4}, simulation results for the atomic physics experiment are shown in dots and calculation from Eq.~(\ref{eq:as2}) is shown as the curve.
(b) The simulated symmetron field profile $\phi-\phi_\text{off}$ is shown in the cylindrical coordinates for $M=10^{20}$~GeV [circled data point in (a)], with the offset $\phi_\text{off}=2\times 10^{32}$~eV.
Note that although the source mass and the wall are unscreened, the difference in the VEVs between regions still causes spatial variations in the field profile and thus symmetron forces.
The small relative variation of $10^{-30}$ is amplified by the huge $\phi_\text{off}$ to generate sizable acceleration.
}
\label{fig:astro}
\end{figure}
Constraints on symmetron theory have also been established through astrophysical observations.
By comparing distances of remote stars estimated from methods either sensitive or insensitive to the screening mechanism, bounds on the screening effect and thus on the theories are obtained~\cite{burrage2018tests,sakstein2017tests,jain2013astrophysical,vikram2014astrophysical}.
Prior to this work, the constraints from analyzing laboratory experiments in atomic physics are far away from the astrophysical bounds in the parameter space.
Thanks to the discovery of Region~{\large \textcircled{\small 4}} from laboratory symmetron acceleration measurements, we are able to extend the atomic physics constraints by more than 20 orders of magnitude to overlap with those from the astrophysical observations where $M$ is of the magnitude of $10^{15}$~GeV or higher.
Figure~\ref{fig:astro} shows simulation results (dots) for $\mu=10^{-4}$~eV, with the range of $M$ extended to $10^{30}$~GeV.
Also shown in the figure is the bound calculated based on Eq.~(\ref{eq:as2}).
Atomic exclusion of $\lambda$ is the shaded area under the bound.
The astrophysical constraint, shown in yellow, is reproduced from Ref.~\cite{burrage2018tests}.
As shown in Fig.~\ref{fig:astro}, not only the predicted dependence of Eq.~(\ref{eq:as2}) but also simulation results make direct comparison in the parameter regions relevant to astrophysics.

One may argue that since all materials in the experiment are unscreened  ($\rho<\rho_*$) in Region~{\large \textcircled{\small 4}} (see Sec.~\ref{iiid}), a detailed modeling of the environment including the optical table, the floor and ceiling of the building is required for a reliable simulation.
In fact, the Compton wavelength of the symmetron field at $\mu=10^{-4}$~eV is $m_s^{-1}=2^{-1/2}\mu^{-1}(1-\rho/\rho_*)^{-1/2}\simeq2^{-1/2}\mu^{-1}\simeq1.4$~mm, which is short compared to distances of surrounding objects to the experiment.
Thus, the symmetron field in air reaches the VEV around the experiment, effectively isolates the experiment from its environment.
This phenomenon is evident in Fig.~\ref{fig:astro}(b), where the symmetron field settles to $\phi_e$ of the air, in a short distance.
The validity of the simulation result is thus based on a reasonable assumption that the experimental apparatus of Ref.~\cite{BerkeleyNaturePhys} is mostly surrounded by a layer air of at least 2~cm, and that supporting structures to the apparatus would only mildly modify the field inside the vacuum chamber.

The astrophysical constraint is about 10 orders of magnitude more stringent than the current atomic physics constraint for $M>10^{15}$~GeV, though the dependence on $\mu$ is not elaborated.
The similarity of $M$-$\lambda$ dependence is a validation of both approaches.
While new astrophysical data or analysis could improve the constraint further, specifically designed atom interferometer experiments may provide definite enhancements in near future.

It is also interesting to note that in Region~{\large \textcircled{\small 4}} all materials are unscreened, so that the advantage of using atoms may not be significant as in other regions.
Reanalysis of results using macroscopic test objects, as was done in Ref.~\cite{upadhye2013symmetron}, may provide a better bound than atomic tests here after addressing potential practical systematics.

\section{Discussion and Conclusions}\label{secIV}
For a given limit on fifth force placed by atomic test particles, the bounds of symmetron parameters are determined by the vacuum density ($M_\text{I}$), the nuclear  properties ($M_\text{II}$), the ambient air density ($M_\text{III}^*$), the size of vacuum chamber ($\mu_w$), the characteristic length ($r_*$), and weakly on the property of the material and configuration ($M_\text{III}$) (Fig.~\ref{fig:bounds}).
We numerically verify the validity of the dependences even in the situation where a complicated source mass structure is used.
To have the most sensitivity to the symmetron force, the characteristics $M_\text{I},M_\text{II},M_\text{III}^*,\mu_w,r_*$ should be as small as feasible, while $M_\text{III}$ should be as large as possible to maintain the $\lambda\propto M^{-2}$ roll-off before $M>M_\text{III}$.
In the following, we will discuss how an ideal atomic test can be constructed and the limitations.
\begin{itemize}[leftmargin=*]
\item $M_\text{I}\propto\sqrt{\rho_o}$, where the number density of hydrogen molecules is $2\times10^7$~/cm$^3$ in a typical laboratory UHV system of about $6\times 10^{-10}$~torr.
In the interplanetary space, on the other hand, the particle density due to the solar wind is less than 10~/cm$^3$.
Thus, an atom interferometer experiment conducted in the direct space vacuum at places such as the cislunar space where a gateway will be deployed will have an improvement of 3 orders of magnitude~\cite{LOPG,LOPG2,LOPG3}.

\item $M_\text{II}\propto\sqrt{\rho_N}R_N\propto A^{1/3}$, where $A$ is the mass number of an atom.
Using lower mass atoms is thus advantageous for reducing $M_\text{II}$.
However, the improvement is marginal: it is only a factor of 5 from $^{133}$Cs to $^1$H.
Considering the maturity of atom interferometry using light atoms, technical challenges out-weigh the benefit.

\item $M_\text{III}\propto\sqrt{\rho_w}$ depends on the density of the source mass and the wall, as well as the thickness of the materials.
It is advantageous to use high density material and thick structures to increase $M_\text{III}$.
The choice of material density is very limited, and the contribution to $M_\text{III}$ will be less than a factor of 5.
Similarly, increasing the wall thickness will help, but in the limit of infinitely thick walls $M_\text{III}$ will still be dominated by the wall density $\rho_w$.
The downside of dense material, as discussed in Ref.~\cite{chiow2018multiloop}, is the gravity of the material.
Ten times denser material will require ten times more stringent tolerance in source mass dimensions, and at the same time the gain in sensitivity may not warrant the technical effort.
The material choice will be a tradeoff between sensitivity and systematics, and the evaluation will be based on detailed analysis and specific science objective for each experiment.

\item $M_\text{III}^*\propto\sqrt{\rho_\text{air}}$, limited by the ambient air density, and can be easily improved by several orders of magnitude by enclosing the vacuum chamber in a large evacuated container.
Performing the experiment directly in outer space vacuum will also eliminate the boundary set by $M_\text{III}^*$ in Fig.~\ref{fig:bounds} and only that of $M_\text{I}$ will remain.

\item $\mu_w\propto 1/R_w$, limited by the size of the vacuum container.
It can be reduced significantly by increasing the size of vacuum chamber, which however is not desirable for space missions.
On the other hand, an experiment conducted directly in the open-space vacuum will completely remove the constraint set by $\mu_w$.

\item $r_*$ is determined by the distance between atoms and the source mass.
A near surface atom interferometer experiment with sub $\mu$m distances will boost the sensitivity by 4 orders of magnitude from the proposed experiment of $r_*\simeq0.5$~cm~\cite{chiow2018multiloop}.
There are technical issues to be addressed before embracing this advantage.
The size of a dilute ultra-cold cloud of $10^6$ atoms is on the order of mm; smaller than that the mutual interactions between atoms may cause phase shifts and systematics.
Moreover, material surface effects such as the Casimir effect, the van der Waals force or patch charges will become significant in the $\mu$m range and below.
More fundamentally, the laser beams that drive atom interferometer operations need to be thicker than the cloud size for intensity uniformity.
Clipping and scattering from the source mass may hinder the performance of atom interferometers.
Thus, a few-mm distance as proposed in Ref.~\cite{chiow2018multiloop} is a good compromise between atom interferometer performance and symmetron sensitivity.
\end{itemize}

To summarize, we established a closed-form expression of symmetron acceleration experienced by atomic test particles.
The closed-form expression employs 5 physical parameters, most of which can be estimated fairly well without resorting to simulation.
A recipe for obtaining the parameters is introduced and validated in two distinct configurations using a software package that can handle more than 100 orders of magnitude in parameters.
We utilized the closed-form expression and the simulation package to extend the bound of symmetron parameters based on the published results of a laboratory experiment.
The excluded region is expanded by more than 20 orders of magnitude in $M$ and 10 orders of magnitude in $\mu$ from those in publications.
As a result, a direct connection with constraints by set astrophysical observations is made for the first time.
Further, an analysis on the proposed atomic tests of chameleon force in space~\cite{chiow2018multiloop} was conducted for symmetron force.
We found that the it would improve the atomic constraints by at least 2 orders of magnitude, and by 5 orders of magnitude near the peak of the sensitivity.
We also discussed potential enhancement based on the closed-form expression of symmetron acceleration.
It is promising of gaining few more orders of magnitude in several fronts of the exclusion plot by performing such an atomic test in open-space vacuum in outer space.

\begin{acknowledgments}
The authors would like to acknowledge valuable discussions with Jason Rhodes, Olivier Dor{\'e}, Phil Bull, J{\'e}r{\^o}me Gleyzes, Jeffrey Jewell, Eric Huff, and Holger M{\"u}ller.
This work was carried out at the Jet Propulsion Laboratory, California Institute of Technology, under a contract with the National Aeronautics and Space Administration. \copyright \hspace{1pt} 2019. California Institute of Technology. Government sponsorship acknowledged.
\end{acknowledgments}

\bibliographystyle{unsrt}
\bibliography{DESymmetron}

\end{document}